\documentclass[fleqn,twoside]{article}
\usepackage[headings]{espcrc2}

\usepackage{graphicx}
\usepackage[figuresright]{rotating}

\newcommand{\AmS}{{\protect\the\textfont2
  A\kern-.1667em\lower.5ex\hbox{M}\kern-.125emS}}

\title{Redundant failures of the dip model of
the extragalactic cosmic radiation }

\author{Antonio Codino\address[PG]{INFN and Dipartimento di Fisica dell'Universit\`a  di Perugia,
Italy.}}

\begin{document}

\begin{abstract}

\par\parskip=1.truecm

The proton flux  and the chemical composition of the cosmic
radiation measured, respectively,  by the Kascade and Auger
experiments entail radical changes in Cosmic Ray Physics.  A large
discrepancy  emerges  by comparing the proton flux predicted by the
dip model and that measured by Kascade in the critical energy
interval $5\times10^{16}$-$10^{17}$ eV.
 It is mentioned and substantiated that the  proton flux measurements
 of the Kascade experiment are  consistent with other pertinent empirical
 observations. It is shown that the
chemical composition measured by Auger by two independent
procedures,  using the mean depth reached by cosmic nuclei in giant
air cascades, is incompatible with that predicted by the dip model.
 \newline \quad A notable consequence suggested here based on the
 failures of the dip model
  is that the spectral index softening of the primary cosmic radiation above
 $6\times 10^{19}$ eV observed by HiRes and Auger experiments,
  is not due to the
 extragalactic cosmological protons suffering energy losses in the intergalactic space
 via the reactions,  $p$ $\gamma$  $\rightarrow$  $\pi^0$
 $p$, $\pi^+$ $n$,
 but to some physical phenomena occurring in the cosmic
 vicinity.

\end{abstract}
\vspace{-0.5cm}

\maketitle


\section{Introduction}
An abrupt progress has been recently occurred in Cosmic Ray Physics,
     which in many respects is a revolutionary change of the current
     notions of the discipline, due to
some measurements  of the Auger and Kascade Collaborations.

The Auger experiment has determined the chemical composition of the
cosmic radiation by measuring the mean depth reached by cosmic
nuclei in giant terrestrial cascades in air
\cite{{augerloga},{sigmaauger2009}} or the equivalent variable,
$X_{max}$. By two independent methods it has been established that
 in the energy interval $4\times10^{17}$-$5\times10^{19}$ eV the
cosmic radiation attaining the solar cavity  consists predominantly
of intermediate nuclei,  and not of pure protons. Figure 1 reports
the $X_{max}$ versus energy measured by the Auger experiment (red
dots) \cite{augerloga}.

\par The Auger apparatus determines  $X_{max}$  by
fluorescent light released by nuclei penetrating the air. The
longitudinal light profile recorded by the instrument is
interpolated by an appropriate function, $f_{LP}$, which has a
characteristic width denoted $\sigma({X_{max}})$.  The measurement
of $\sigma({X_{max}})$ at a given energy for a number of atmospheric
cascades determines the average value of the chemical composition of
the cosmic radiation, which is a second method, besides $X_{max}$.
Figure 2 shows $\sigma({X_{max}})$ versus energy (black dots)
measured by Auger \cite{sigmaauger2009} along with its theoretical
estimates of $\sigma({X_{max}})$ for iron nuclei (turquoise lower
band) and for protons (turquoise upper band).

In a series of measurements the Kascade experiment has determined
that the proton flux in the energy interval
$5\times10^{16}$-$10^{17}$ eV  is $(1-2) \times 10^{15}$
particles/($m^2$ sr s $eV^{1.5}$)
\cite{{kaskaqgsjet},{kaskasibyll}}. The proton flux has been
measured by two methods hereafter referred to as QGSjet and Sibyll
algorithms. Figure 3 reports the proton energy spectrum measured by
Kascade using the QGSjet algorithm \cite{kaskaqgsjet}. Data from the
Sibyll algorithm \cite{kaskasibyll} are equivalent for the purpose
of this study and omitted for brevity.


\begin{figure}[htb]
\vspace{-0.3cm}
\includegraphics [angle=0,width=8cm,height=8cm] {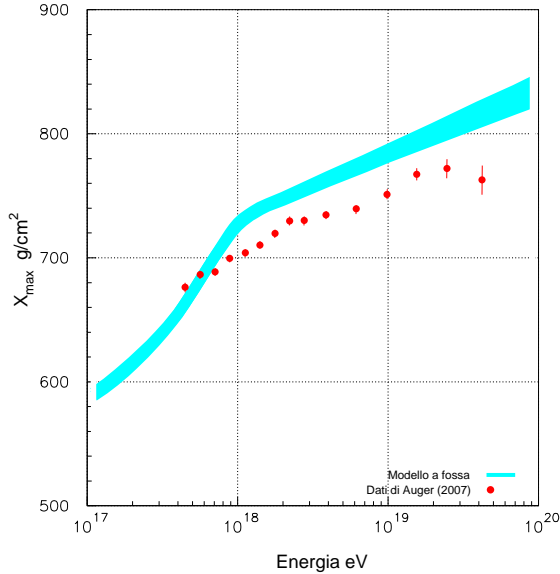}
\vspace{-1.5cm} \caption{ Measurements of  $X_{max}$ by Auger (red
dots) compared with the corresponding $X_{max}$ derived from the dip
model (turquoise band). The lower and upper theoretical values of
$X_{max}$  of the dip model
 at a given energy correspond to different theoretical models of
hadronic interactions in air.} \label{fig:largenenough}
\vspace{-0.3cm}
\end{figure}

It is the aim of this $\it {Letter}$ to show that the three
independent quoted measurements  are incompatible with the
corresponding predictions of the  dip model
\cite{{modelloafossa1},{fossa2}}.

   This $\it {Letter}$ benefits of  a critical examination  of the HiRes data
on $X_{max}$ made in another study \cite{codplochemical}.  Figure 4
shows that above  $10^{17}$ eV the HiRes data  \cite{belzsalina} on
$X_{max}$ converted into $<$$ln(A)$$>$ differ systematically
 from those of the Volcano Ranch, Yakutsk, Akeno, Agasa,
Haverah Park, Fly's Eye and Auger experiment \cite{codplochemical}.
Figure 5 shows that the $\sigma({X_{max}})$ profile measured by
HiRes  \cite{sokolskinorvegia}   also differs from that measured by
Auger \cite{sigmaauger2009} .


\begin{figure}[htb]
\vspace{-0.3cm}
\includegraphics [angle=0,width=8cm,height=8cm] {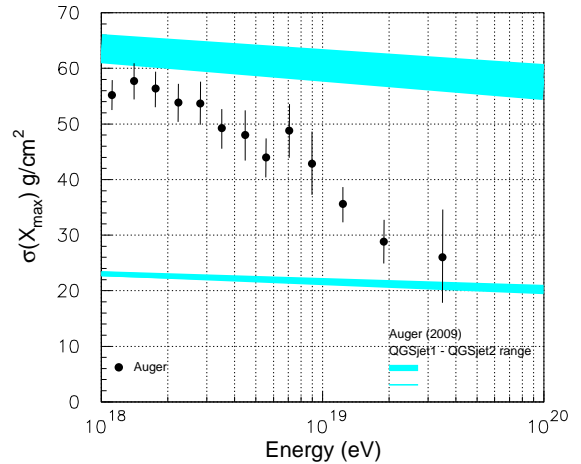}
\vspace{-1.5cm} \caption{Measurements of  $\sigma({X_{max}})$  by
Auger (black dots) and the related theoretical estimates
\cite{sigmaauger2009} for a cosmic radiation consisting of protons
only (upper colored band) and Fe nuclei only (lower colored band).}
\label{fig:largenenough} \vspace{-0.4cm}
\end{figure}


\begin{figure}[htb]
\vspace{-0.3cm}
\includegraphics [angle=0,width=8cm,height=8cm] {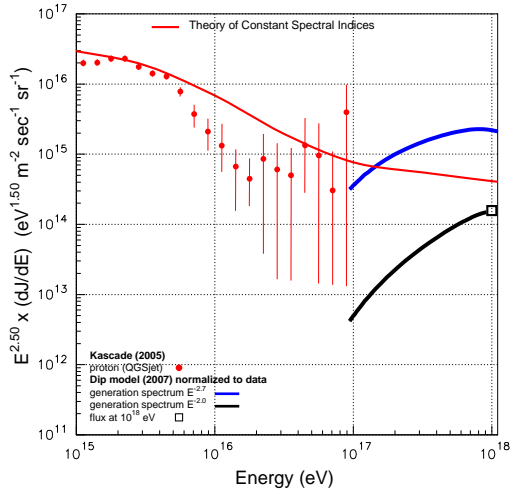}
\vspace{-1.5cm} \caption{Extragalactic proton flux at $10^{18}$ eV
according to the dip model \cite{aloisiofigure2} (black square) with
the standard index  $\gamma_s$= $2$  and its extrapolation (black
curve) down to $10^{17}$ eV calculated in this study. The proton
spectrum of the same model with $\gamma_s$= $2.7$ (blue curve)
normalized $\it {ad}$ $\it {hoc}$ to some experimental data
\cite{aloisiofigure2}, that measured by Kascade (red dots)
\cite{kaskaqgsjet}, and that predicted by the $\it {Theory}$ $\it
{of}$ $\it {Constant}$ $\it {Spectral}$ $\it {Index}$ (red curve)
are also displayed for comparison \cite{flussimisurati}.}
\label{fig:largenenough} \vspace{-0.4cm}
\end{figure}

\section{Measurements of  $X_{max}$ and $\sigma({X_{max}})$
and related predictions of the the dip model.}

The dip model  proposed in 1988 states that  extragalactic protons
originated at cosmological distances can reach copiously the solar
cavity. They interact with cosmic fossile photons $\gamma$ with
energies centered around  $6\times10^{-4}$ eV and density of $420$
$particles$/$cm^3$, via the reaction: p $\gamma$ $\rightarrow$ $e^-$
$e^+$ $p$, where $p$ denotes extragalactic proton and $e^-$, $e^+$
electron pairs. The kinematical threshold of this reaction is at
$4\times10^{17}$ eV, a basilar reference energy of the dip model.
Extragalactic protons would suffer energy losses in the
intergalactic space via the quoted reaction generating a depression
in the original unperturbed spectrum released by the extragalactic
accelerator which has constant proton index $\gamma_s$ in the range
2.0-2.7.

The dip model does not specify a precise acceleration mechanism, the
exact sites where the accelerators operate, the spectral indices of
the cosmic ions at the cosmic-ray sources (they are free parameters
condensed in a single one $\gamma_s$), the ion filtering at the
injection to the accelerators at the low energy and other
parameters. It preassumes that in the intergalactic proton
displacement, besides the expansion of the universe,   no major
processes other than the reaction, p $\gamma$ $\rightarrow$ $e^-$
$e^+$ $p$,  intervene. While most of the unknowns of the dip model
enumerated above are not surprising in the present status of the
discipline, the intergalactic proton displacement affected only by
the reaction, p $\gamma$ $\rightarrow$  $e^-$ $e^+$ $p$  (for
instance, de-acceleration or re-acceleration processes with uneven
magnitudes may take place) and a constant instead of a variable
$\gamma_s$ are rather fragile hypotheses.


\begin{figure}[htb]
\vspace{-0.3cm}
\includegraphics [angle=0,width=8cm,height=8cm] {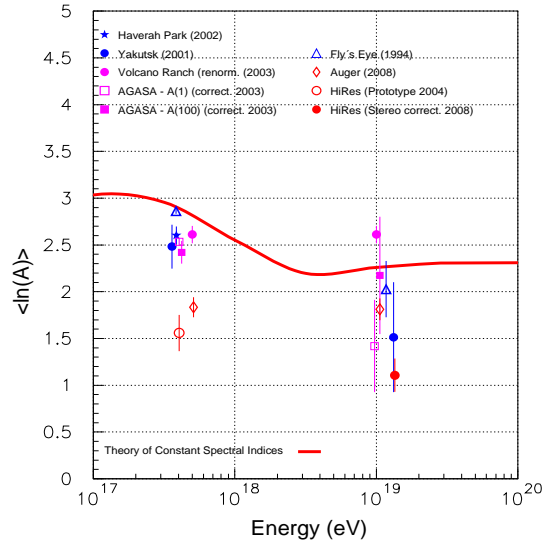}
\vspace{-1.5cm} \caption{Mean values of the chemical composition
expressed in terms of $<$$ln(A)$$>$ at the two arbitrary energy
bands
 of $(2-7)\times10^{17}$ eV and $(0.6-2.1)\times10^{19}$ eV,
for 9 experiments according to a recent homogeneous evaluation
\cite{codplochemical} (see references therein). The HiRes data
occupy the extreme lowest values of $<$$ln(A)$$>$ out of all
measurements. The $<$$ln(A)$$>$  predicted by the  $\it {Theory}$
$\it {of}$ $\it {Constant}$ $\it {Spectral}$ $\it {Index}$  (red
curve) \cite{codplochemical}  is shown as a useful guide to the
data.} \label{fig:largenenough} \vspace{-0.4cm}
\end{figure}


\begin{figure}[htb]
\vspace{-0.3cm}
\includegraphics [angle=0,width=8cm,height=8cm] {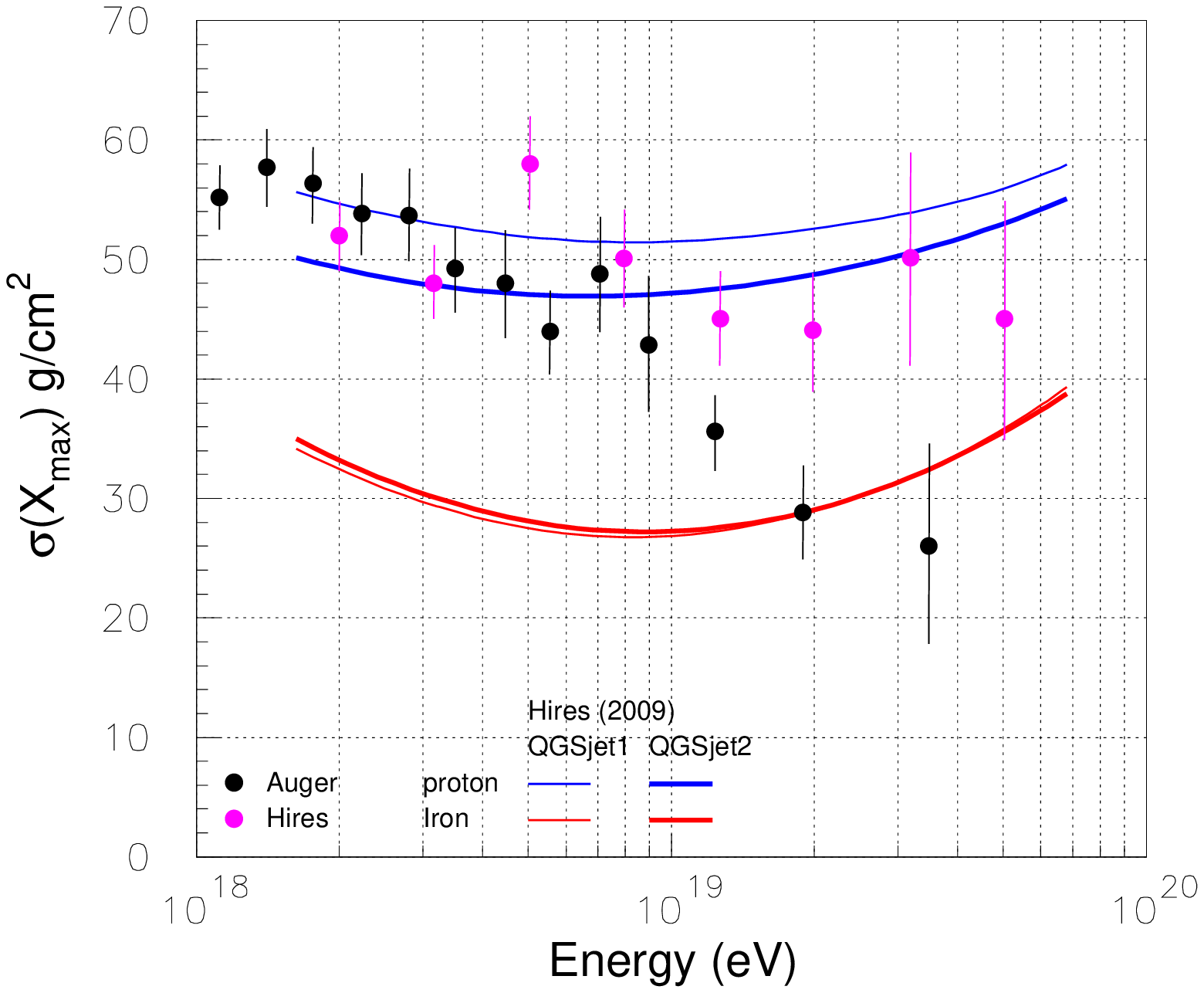}
\vspace{-1.5cm} \caption{Measurements of $\sigma({X_{max}})$ of the
Auger \cite{sigmaauger2009} and HiRes experiments
\cite{sokolskinorvegia} compared with the corresponding Monte Carlo
simulations  of nuclear interactions, by the QGSjet-01 and QGSjet-02
codes, assuming that all cosmic rays are protons (upper blue curves)
or Fe nuclei (lower red curves) according to HiRes
\cite{sokolskinorvegia} . } \label{fig:largenenough} \vspace{-0.4cm}
\end{figure}

The chemical composition of the dip model has been calculated and
converted by others  \cite{aloisiofigure2}  into the corresponding
$X^{th}_{max}$ versus energy using three different hadronic codes
denoted QGSjet, QGSjet-2 and Sibyll \cite{aloisiofigure2}. The
minimum and the maximum values of the $X^{th}_{max}$ of the dip
model at a given energy are shown in figure 1 defining the colored
turquoise band. It turns out that the Auger data are below the
minimum values of the predicted $X^{th}_{max}$  profile.

If the cosmic radiation consisted of Fe nuclei only, the
$\sigma({X^{Fe}_{max}})$ would occupy the turquoise lower band in
figure 2, which spans from 23 $g$/$cm^2$ at $10^{18}$ eV down to 20
$g$/$cm^2$ at $10^{20}$ eV \cite{sigmaauger2009}.   Similarly, the
$\sigma({X^{H}_{max}})$ of pure proton cosmic rays would occupy  the
upper turquoise band  in Figure 2, according to the QGSjet code
adopted by Auger \cite{sigmaauger2009}. The simulated
$\sigma({X^{Fe}_{max}})$ profile is less sensitive to the QGSjet
code (quite narrow band), than the $\sigma({X^{H}_{max}})$ profile
(large band). According to the 13 data points on $\sigma({X_{max}})$
measured by Auger, the primary cosmic radiation around
$3\times10^{19}$ eV consists abundantly of intermediate and heavy
nuclei, since the computed $\sigma({X^{Fe}_{max}})$ and the observed
$\sigma({X_{max}})$ almost joins.

Taking into account the dependence of $\sigma({X^{A}_{max}})$ on the
mass $A$ of the cosmic nucleus \cite{sigmaauger2009}, any ion
abundances at a given energy can be converted into
$\sigma({X_{max}})$. It results that the Auger data in Figure 2
above $10^{19}$  eV exclude a proton dominance  in the cosmic
radiation as foreseen by the dip model (some 80-90 $\%$ as stated in
ref. \cite{aloisiofigure2}).

\section{Proton flux measurements in Kascade and the
related dip model predictions}

The proton flux  at $10^{18}$ eV derived from the dip model in its
standard form and
 normalization with $\gamma_s$=2.0 \cite{aloisiofigure2} is
 $1.58\times10^{14}$ particles/($m^2$ sr s $eV^{1.5}$)
(black square in fig. 3). It results that  the discrepancy between
measured and theoretical flux (black square and extrapolated black
curve) grazes more than two orders of magnitude at $10^{17}$ eV
since the predicted extragalactic proton spectrum below $10^{18}$ eV
has to descend whatsoever with decreasing energy  from the value of
$1.58\times10^{14}$ particles/($m^2$ sr s $eV^{1.5}$) at $10^{18}$
eV.

The empirical foundation and solidity of this conclusion is examined
 in detail elsewhere \cite{flussimisurati}: proton fluxes in the  Kascade data
 samples
 in the interval $5\times10^{16}$
-$10^{17}$ eV exceed those obtained by  QGSjet and Sibyll algorithms
because of some contamination
 of the helium sample by protons (see Section 7 of ref. \cite{flussimisurati}).
Enhanced proton data samples augment the  flux and decrease error
magnitudes in Figure 3.

 Notice that the silouhette of the proton
spectrum according to the dip model \cite{aloisiofigure2}
 between  $10^{17}$ and $10^{18}$ eV with $\gamma_s$=2.7
shown in Figure 3 (blue curve) is $\it {arbitrarily}$  and $\it
{ad}$ $\it {hoc}$  normalized to some experimental data in the band
$(5-8)\times10^{17}$ eV.

\section{Data and cascade simulation codes in HiRes}

Figure 5 shows the estimated theoretical profiles of
$\sigma({X^{H}_{max}})$ and $\sigma({X^{Fe}_{max}})$  according to
the HiRes experiment which adopts the hadronic codes QGSjet-1 and
QGSjet-2 \cite{sokolskinorvegia}. The first seven data points from
Auger (black dots) in the interval $10^{18}$-$5\times10^{18}$ eV
would fall above the theoretical $\sigma({X^{H}_{max}})$ profile
obtained by  the QGSjet-2 code. Therefore, eight Auger data points
out of 13 would become unphysical (e.g. cosmic particles lighter
than protons). Similarly, the last data point (Auger) is positioned
below the theoretical profile $\sigma({X^{Fe}_{max}})$ suggesting
that hyperheavy cosmic nuclei ($A$ $>$ $56$)  dominate the cosmic
radiation above $3\times10^{19}$ eV, or more plausibly, again, an
unphysical condition develops. It may not be surprising that, in two
independent areas of comparison (on $X_{max}$ and
$\sigma({X_{max}})$, the outcomes of the HiRes Collaboration might
disagree with all other experiments. What is both surprising and
embarrassing  is that the hadronic codes to simulate nuclear
interactions in air, in HiRes, over many years, generate more
protons and less heavy nuclei than hadronic codes adopted in all
other experiments.  Figure 5 vividly demonstrates it in a recent
example \cite{sokolskinorvegia}.

\section{Conclusions}

The fundamental tenet of the dip model  is that protons observed in
the solar cavity above $4\times 10^{17}$ eV have a cosmological
origin
 and outnumber any other ion fractions. Since the dip model
predicts unreal properties of the cosmic radiation, the $Logic$
dictates, out of a few alternatives, that spatial sources of the
cosmic-ray protons are in the cosmic vicinity ($\it {Galaxy}$ or
$\it{Local}$ $\it{Group}$  or $\it{Local}$ $\it{Supercluster}$
  $\it{of}$
$\it{Galaxies}$). In this circumstance: how could a softening of the
spectral index from 2.6 to about 3.5 take place in the interval
$5\times10^{19}$-$3\times10^{20}$ eV, if the extragalactic
cosmological protons below $5\times10^{19}$ eV do not reach the
solar cavity ?

As the reaction $p$ $\gamma$ $\rightarrow$  $e^-$ $e^+$ $p$ does not
alter any cosmic-ray spectrum in the solar cavity (first
alternative), the companion reactions $p$ $\gamma$ $\rightarrow$
$\pi^0$ $p$, $\pi^+$ $n$, etc.  can not cause the index softening
above $5\times10^{19}$ eV just because the extragalactic
cosmological protons are missing.

The $ad$ $hoc$ hypothesis that a cosmological extragalactic
component would reach the solar cavity in the interval
$6\times10^{19}$-$3\times10^{20}$ eV but not below this energy band
(second alternative) confront with three barriers: (1) protons
suffering the energy losses via $p$ $\gamma$ $\rightarrow$ $\pi^0$
$p$ do not disappear but they accumulate below $6\times10^{19}$ eV
corrugating the spectrum; the dip model, as an example,  attempts to
calculate this effect with a simplified calculation scheme. (2) Any
extragalactic accelerators, operating in the specific interval
$6\times10^{19}$-$3\times10^{20}$, and eventually above this maximum
energy, would intrinsically  generate enough spill-over particles in
the low energy edge, below $6\times10^{19}$, which plausibly
corrugate the galactic spectrum. (3) The rising dominance of heavy
ions with energy discovered by the Auger Collaboration (see figures
1 and 2) in the band $10^{19}$-$4.12\times10^{19}$ eV does not
harmonize, for a number of reasons, with the aforementioned index
softening.

The common value of 2.6-2.8 of the spectral index of the cosmic
radiation close and below $6\times10^{19}$ eV observed in all
experiments (HiRes, Auger, Agasa, Yakutsk), disfavors and belittles
the second alternative.

According to the present investigation, the softening of the
spectral index of the cosmic radiation above $6\times10^{19}$ eV is
not due to any extragalactic cosmological protons.  This conclusion
might represent the most notable result rooted to the reaction $p$
$\gamma$ $\rightarrow$  $e^-$ $e^+$ $p$ as conceived and framed in
the dip model.


\end{document}